%Paper: astro-ph/9304005
%From: BEST%IASSNS.BITNET@PUCC.PRINCETON.EDU
%Date: Tue, 6 Apr 93 10:58 EST

\font\titlefnt=cmbx10 scaled \magstep2
\font\sc=cmbx12
\font\secfnt=cmbx12
\font\subfnt=cmsl12

\def\ctr#1{\hfill#1\hfill}                            %put #1 in center of box
                                   %put #1 at right of box
                                  %put #1 at left of box

\def\linebreak{\hfil\break}

 %restarts numbering after
                                                   % \nopagenumbers
\catcode`\@=11 %insert \rm to ensure correct font in footnotes
\def\vfootnote#1{\insert\footins\bgroup\rm
  \interlinepenalty\interfootnotelinepenalty
  \splittopskip\ht\strutbox % top baseline for broken footnotes
  \splitmaxdepth\dp\strutbox \floatingpenalty=20000
  \leftskip=0pt \rightskip=0pt \spaceskip=0pt \xspaceskip=0pt
  \textindent{#1}\footstrut\futurelet\next\fo@t}
\catcode`\@=12
\newcount\secno
\newcount\sequationno
\newcount\equationno
\def\sect#1\par{
  \global\advance\secno by1
  \global\sequationno=0  %reset  section equation counter to 0
  \vskip0pt plus.2\vsize\penalty-100 %for section headings
  \vskip0pt plus-.2\vsize\bigskip\vskip\parskip
  \message{#1}
  \centerline{\secfnt #1}
  \nobreak\medskip}
\def\sectnonumber#1\par{
  \global\sequationno=0  %reset equation counter to 0
  \vskip0pt plus.2\vsize\penalty-100 %for section headings
  \vskip0pt plus-.2\vsize\bigskip\vskip\parskip
  \message{#1}
  \centerline{\secfnt #1}
  \nobreak\medskip}
\def\subsect#1\par{\vskip0pt plus.2\vsize\penalty-50 %subsections
  \vskip0pt plus-.2\vsize\bigskip\vskip\parskip
  \message{#1}
  \centerline{\subfnt#1}
  \nobreak\medskip}
\def\appendix#1\par{\sectnonumber#1\par}

\def\title#1{\vskip 24pt plus 12pt minus 12pt\tabskip0pt plus 1000pt
   \halign to \hsize{\titlefnt\ctr{##}\cr#1\crcr}\medskip}

\def\author#1{\bigskip\tabskip0pt plus 1000pt \halign to
\hsize{\sc\ctr{##}\cr#1
   \crcr}}

\def\affil#1{\smallskip\tabskip0pt plus 1000pt
  \halign to \hsize{\rm\ctr{##}\cr#1\crcr}}

\def\date#1{\bigskip\centerline{\rm #1}\bigskip}

\def\abstract{
   \sectnonumber Abstract\par \leftskip=20pt\rightskip=20pt}
\def\endabstract{\par\leftskip=0pt\rightskip=0pt}

\def\ack#1{\bigbreak\centerline{\sc Acknowledgements}\nobreak\medskip#1}

\def\Refs{\sectnonumber References\par\frenchspacing}
\def\refindent{\par\penalty-100\noindent
               \hangindent 20pt\hangafter=1\null}
\def\ref#1#2#3#4{\refindent#2, {\it #1\/,\ }{\bf#3}, #4.}
\def\refbook#1{\refindent#1}
\def\preprint#1#2#3{\refindent#1, #2, {\it #3 preprint}.}

\def\bysame{\hbox to 50pt{\leaders\hrule height 2.4pt depth -2pt\hfill .\ }}

\def\apjstyle{
  \def\aa{{Astr.\ Ap.}}
  \def\aj{{A.J.}}
  \def\annrev{{Ann.\ Rev.\ Astr.\ Ap.}}
  \def\apj{{Ap.\ J.}}
  \def\apjl{{Ap.\ J.\ (Letters)}}
  \def\apjs{{Ap.\ J.\ Suppl.}}
  \def\mn{{M.N.R.A.S.}}
  \def\nat{{Nature}}
  \def\pasp{{Pub.\ A.S.P.}}
  \def\rmp{{Rev.\ Mod.\ Phys.}}
}

\def\ceqno{\global\advance\equationno by1
           \eqno{(\the\equationno)}}
\def\ceqnoa{\global\advance\equationno by1
            \eqno{(\the\equationno a)}}
\def\ceqnob{\eqno{(\the\equationno b)}}
\def\ceqnoc{\eqno{(\the\equationno c)}}
\def\ceqnod{\eqno{(\the\equationno d)}}
\def\ceqnoe{\eqno{(\the\equationno e)}}

\def\ceqalign{\global\advance\equationno by1
              {(\the\equationno)}}
\def\ceqaligna{\global\advance\equationno by1
               {(\the\equationno a)}}
\def\ceqalignb{{(\the\equationno b)}}
\def\ceqalignc{{(\the\equationno c)}}
\def\ceqalignd{{(\the\equationno d)}}
\def\ceqaligne{{(\the\equationno e)}}

\def\seqno{\global\advance\sequationno by1
           \eqno{(\the\secno.\the\sequationno)}}
\def\seqnoa{\global\advance\sequationno by1
            \eqno{(\the\secno.\the\sequationno a)}}
\def\seqnob{\eqno{(\the\secno.\the\sequationno b)}}
\def\seqnoc{\eqno{(\the\secno.\the\sequationno c)}}
\def\seqnod{\eqno{(\the\secno.\the\sequationno d)}}
\def\seqnoe{\eqno{(\the\secno.\the\sequationno e)}}

\def\seqalign{\global\advance\sequationno by1
              {(\the\secno.\the\sequationno)}}
\def\seqaligna{\global\advance\sequationno by1
               {(\the\secno.\the\sequationno a)}}
\def\seqalignb{{(\the\secno.\the\sequationno b)}}
\def\seqalignc{{(\the\secno.\the\sequationno c)}}
\def\seqalignd{{(\the\secno.\the\sequationno d)}}
\def\seqaligne{{(\the\secno.\the\sequationno e)}}

\def\apeqno{\global\advance\sequationno by1
             \eqno{(\rm A.\the\sequationno)}}
\def\bpeqno{\global\advance\sequationno by1
             \eqno{(\rm B.\the\sequationno)}}
\def\cpeqno{\global\advance\sequationno by1
             \eqno{(\rm C.\the\sequationno)}}

\def\apeqalign{\global\advance\sequationno by1
               {(\rm A.\the\sequationno)}}
\def\bpeqalign{\global\advance\sequationno by1
               {(\rm B.\the\sequationno)}}
\def\cpeqalign{\global\advance\sequationno by1
               {(\rm C.\the\sequationno)}}

\equationno=0          %default starting eqn no. -1
\sequationno=0          %default starting eqn no. -1
\secno=0               %default starting section no. -1
\apjstyle              %default to Ap. J.-style reference abbrevs.

\def\sun{{_{\odot}}}                                    % Sun
                                     % Earth
 \mathcode`*="002A              %makes * ok in math
\def\simless{\mathbin{\lower 3pt\hbox
   {$\rlap{\raise 5pt\hbox{$\char'074$}}\mathchar"7218$}}} %< or of order
\def\simgreat{\mathbin{\lower 3pt\hbox
   {$\rlap{\raise 5pt\hbox{$\char'076$}}\mathchar"7218$}}} %> or of order
\def\doublearrowfill{$\mathord\leftarrow\mkern-6mu%
  \cleaders\hbox{$\mkern-2mu\mathord-\mkern-2mu$}\hfill
  \mkern-6mu\mathord\rightarrow$}
\def\overdoublearrow#1{\vbox{\ialign{##\crcr
      \doublearrowfill\crcr\noalign{\kern-1pt\nointerlineskip}
      $\hfil\displaystyle{#1}\hfil$\crcr}}}
\def\th{\hbox{.\kern.6em \hbox{.}\kern-.69em \raise1.3ex \hbox{.}}\quad}%
     %3 dots therefore symbol
\def\Islash{{I \kern-4pt \raise+1.5pt \hbox{-}}}
\def\boxit#1{\vbox{\hrule\hbox
{\vrule\kern3pt\vbox{\kern3pt#1\kern3pt}\kern3pt\vrule}\hrule}}

\def\scin #1 #2 {#1 \times 10^{#2}}

\def\isot #1 #2 {{}^{#2} \hbox{#1}}
\def\isotc #1 #2 #3 {{}_{#2}^{#3} \hbox{#1}}

\def\listitem{\par \hangindent=50pt\hangafter=1 %Bigger indent than item+bullet
     $\ $\hbox to 20pt{\hfil $\bullet$ \hfil}}  %Also subseq. lines in 1 space

\apjstyle
\def\preprint#1{\refindent#1, preprint.}
\newwrite\defs
\openout\defs=\jobname.def
\newcount\eqncnt
\def\+{\number\eqncnt}                   % \+ gives the current eq no.
\def\e{\global\advance\eqncnt by 1 \+}   % \e gives next eq no. without a name
\def\en#1{\e\xdef#1{\+}\sendout#1}       % \en\name names and gives next eq no
\def\sendout#1{\write\defs{\gdef\noexpand#1{#1}}}
\baselineskip = 18 pt
\magnification 1200

\title{Element Diffusion in the Solar Interior}
\bigskip
\author{Anne A. Thoul, John N. Bahcall and Abraham Loeb}
\affil{Institute for Advanced Study\cr Princeton, NJ 08540}
\bigskip

\abstract{
We study the diffusion of helium and other heavy elements
in the solar interior by solving exactly the set of flow equations developed
by Burgers for a multi-component fluid, including the residual
heat-flow terms. No approximation is made concerning the
relative concentrations and no restriction is placed
on the number of elements considered.
We give improved diffusion velocities for hydrogen, helium,
oxygen and iron, in the analytic form derived previously by Bahcall and Loeb.
These expressions for the diffusion velocities are simple to
program in stellar evolution codes
and are expected to be accurate to
$\sim 15\%$.
We find that the inclusion of the residual heat flow
terms leads to an increase in the hydrogen
diffusion velocity.
We compare our numerical
results with those obtained analytically by Bahcall and Loeb
using a simplified treatment, as well as with those derived numerically by
Michaud and Proffitt. We find that for conditions characteristic of the sun,
the results of
Bahcall and Loeb for the hydrogen
diffusion velocity are smaller than our more accurate numerical
results by $\sim 30\%$, except very near the center where the error
becomes larger. The Michaud and Proffitt results differ from
the numerical results derived here by $\simless 15\%$.
Our complete treatment of element diffusion can be
directly incorporated in a standard stellar evolution
code by means of an exportable subroutine, but, for convenience,
we also give simple
analytical fits to our numerical results.}
\endabstract
{\it Subject headings}: diffusion, stars: interiors, stars: abundances, Sun:
int
 erior

\vfill\eject

\sect{1. Introduction}

Precise solar evolution calculations must be carried out to compare
model results with observations of solar neutrino fluxes and of p-mode
oscillation frequencies.
In particular, element diffusion affects the element abundances, the
mean molecular weight, and the
radiative opacity in the core of the sun, and therefore affects the calculated
neutrino fluxes and oscillation frequencies.
The
characteristic time for elements to diffuse a solar radius under solar
conditions is of the order of
$6\times 10^{13}\,{\rm yrs}$, much larger than the age of the sun.
Element diffusion  therefore introduces only a small correction
to standard solar model calculations.
Bahcall and Pinsonneault (1992a,b) showed that helium diffusion
increases the predicted
event rates by  about 11\% in the chlorine solar neutrino experiment, by
3\% in the gallium experiment, and by 12\% in the Kamiokande and SNO
experiments, while increasing the inferred primordial helium abundance by
0.4\% and decreasing the calculated depth of the convection zone by 2\%.
Christensen-Dalsgaard, Proffitt and Thompson (1993) calculated the sound speed
as a function of radius in the solar model and concluded that helium diffusion
causes a significant difference in the computed radial profile of the
sound speed.
Guenther, Pinsonneault, and Bahcall (1993) demonstrated that helium diffusion
has a characteristic which depends upon the degree and frequency of the p-mode
being discussed and which has a typical amplitude of order 1-3 MHz.

Since the effects of diffusion are small, there is in principle no need for
very high
accuracy in its treatment. However, discrepencies appear
between various results in the
literature, depending on the approximations made.
Previous studies of element diffusion in the sun
(Aller and Chapman 1960, Michaud {\it et al.} 1976,
Noerdlinger 1977, 1978,
Cox, Guzik and Kidman 1989,
Paquette {\it et al.} 1986, Bahcall and Loeb 1990,
Proffitt and Michaud 1991, Michaud and Proffitt 1992, Bahcall and
Pinsonneault 1992a,b,
Christensen-Dalsgaard, Proffitt and Thompson 1993, Guenther, Pinsonneault
and Bahcall 1993,
Vauclair and Vauclair 1982 and references therein)
have usually included one or more of the following simplifying
assumptions: neglecting thermal diffusion, or treating it
using a simplified empirical
formula;
neglecting the presence of heavy elements when
calculating helium diffusion; assuming
a negligible helium abundance  when calculating the diffusion of heavier
elements;
adopting a single constant value for all Coulomb logarithms.
In this paper, we provide a simple but
complete treatment of the problem, making none of the above
approximations, and we compare our results with
those obtained under different simplifying assumptions.
In particular, we compare our results with those obtained by
Bahcall and Loeb (1990) (hereafter BL) and those obtained by Michaud
and Proffitt (1992) (hereafter MP). BL made most of the above
simplifying assumptions. In particular, they used empirical results
for the thermal diffusion coefficients and a single value for the Coulomb
logarithms,
equal to 2.2. MP solved the Burgers equations and then represented the
effects of the residual heat flow vectors by an ad-hoc correction to the
results
obtained when neglecting those heat fluxes.
The principal difference between this work and most previous studies
is that we solve the Burgers equations exactly and then represent
the numerical results by simple analytic functions, rather than trying to
obtain analytic solutions by various approximations.

Element diffusion in stars is driven by
pressure gradients (or gravity), temperature gradients, composition
gradients, and radiation pressure\baselineskip=12pt\footnote{$^1$}{In this
work, we ignore the effects of meridional circulation. It has been shown
that meridional circulation velocities are several orders of magnitude smaller
than the diffusion velocities in the solar interior (see, e.g., Michaud and
Vauclair 1991).}\baselineskip=18pt.
Gravity tends to concentrate the heavier elements towards the
center of the star. In a pure hydrogen-helium plasma, helium
diffuses towards the center of the star, while hydrogen diffuses outwards.
As we will show in \S4 (see also Bahcall and Loeb 1990), the local rate
of change of the hydrogen mass
fraction is equal and opposite to the rate of change of the helium mass
fraction.
This follows from the condition of momentum conservation.
The light electrons also tend to rise, but are held back by an electric
field which
counteracts gravity.
Temperature gradients lead to thermal diffusion, which tends to concentrate
more highly charged and more massive species towards the hottest region
of the star, its center.
Concentration gradients oppose the above processes.
Radiation pressure causes negligible diffusion in the solar core
and will be neglected in this paper.

We study the relative diffusion of hydrogen, helium, and heavier
elements, such as oxygen and iron. In contrast to many previous studies, no
approximation is made concerning the relative concentrations
of the various species, and no restriction
is placed
on the number of elements considered.
Our method is therefore applicable to a wide variety of
astrophysical problems, such as the diffusion of elements in white dwarf
envelopes (see, e.g., Fontaine and Michaud 1979, Pelletier {\it et al.} 1986)
and
in globular cluster stars (see, e.g., Chaboyer {\it et al.} 1992). In
this paper, we concentrate
on calculating the diffusion velocities in the temperature and density ranges
relevant to the sun, although our exportable subroutine can be used to
calculate
diffusion velocities in red giants and in white dwarfs.

Burgers (1969) has provided a complete and straightforward  set of equations to
describe  the evolution of a multi-component fluid.
In order to include the effects of
thermal diffusion,
he introduced the so-called ``residual heat flow vectors''.
Here, we will use the Burgers equations,
including the residual heat fluxes, to describe the plasma
in the solar interior.
Even though these equations can
in principle be solved analytically, the algebraic complexity
increases rapidly with the number of species considered. For example,
because of computational limitations, Noerdlinger (1977)
included only three species (hydrogen, helium, and electrons)
and adopted a single constant for all the Coulomb logarithms.
In contrast,
we solve the full set of Burgers equations
numerically, and place no restriction on the number of species.
The Coulomb logarithm is obtained by calculating the
collision integrals
using a pure Coulomb potential with a
long-range cutoff at the Debye length. However, the result obtained
for the Coulomb logarithm is valid only for plasmas
that are sufficiently hot and rarefied, i.e., such that the plasma parameter
$\Lambda$ is much larger than unity. For conditions characteristic of the
solar interior, the Coulomb logarithms are small, and can even become negative
for collisions between heavy elements. For such plasmas, the collision
integrals can be calculated numerically using a screened Debye-Huckel Coulomb
potential. The results can then be fitted to simple analytic functions.
We adopt an expression
for the ``effective'' Coulomb logarithm obtained by
Iben and MacDonald (1985) by fitting numerical results from Fontaine and
Michaud
  (1979b).

It should be relatively easy to incorporate our complete
treatment of element diffusion into any standard solar
evolution code\baselineskip=12pt\footnote{$^2$}{Our
FORTRAN routine will be made available upon request.}\baselineskip=18pt.
However, we have obtained simple analytic fits
to the exact results, which can provide a convenient alternative.
These fits can be expressed as follows:
Following BL's notations (see footnotes 3 and 5) and
using BL's dimensionless variables (see \S2), the mass fraction of
element $s$ satisfies
the equation
$${\partial X_s\over\partial t}=-\,{1\over\rho r^2}\,{\partial\over\partial r}
  [r^2X_sT^{5/2}\xi_s(r)],\eqno(\e)$$
where the partial derivatives are evaluated  in the local rest frame of a mass
shell
in the star, i.e., in Lagrangian coordinates.
The function $\xi_s(r)$ is related to the diffusion velocity $w_s$
of species $s$ through
$$\xi_s(r)=w_s(r)\rho(r)/T^{5/2}(r).\eqno(\e)$$
We have obtained the following results for the diffusion velocities of
hydrogen, oxygen and iron in the solar interior:
$$\xi_s(r)=A_p(s){\partial\ln p\over \partial r}+A_T(s){\partial\ln T
             \over \partial r}+
           A_H(s){\partial\ln C_H\over \partial r}\eqno(\e)$$
with
$$\left\{\eqalign{&A_p(H)=-2.09+3.15\,X-1.07\,X^2,\cr
        &A_T(H)=-2.18+3.12\,X-0.96\,X^2,\cr
        &A_H(H)=-1.51+1.85\,X-0.85\,X^2,\cr}\right.\eqno(\e)$$
for the hydrogen diffusion coefficients,
$$\left\{\eqalign{&A_p(O)=0.15+1.34\,X-0.89\,X^2,\cr
        &A_T(O)=0.53+1.99\,X-0.72\,X^2,\cr
        &A_H(O)=0.08+0.58\,X-0.28\,X^2,\cr}\right.\eqno(\e)$$
for the oxygen diffusion coefficients, and
$$\left\{\eqalign{&A_p(Fe)=0.25+1.31\,X-0.87\,X^2,\cr
        &A_T(Fe)=0.65+1.99\,X-0.75\,X^2,\cr
        &A_H(Fe)=0.09+0.53\,X-0.27\,X^2,\cr}\right.\eqno(\e)$$
for the iron diffusion coefficients.
These fits were obtained by using a constant value for each
Coulomb logarithm, equal to its value at the center of the sun,
and are accurate to better than 15\% for the hydrogen
and oxygen diffusion velocities, and better than
20\% for the iron diffusion velocity.
Of course, the fits can only have a limited domain of applicability,
whereas the numerical routine is completely general.

This paper is organized as follows. In \S2, we introduce the notation and
basic equations.
In \S3, we describe the method of solution.
In \S4, we
give the results for the hydrogen and helium diffusion coefficients, for
a fixed value of the temperature and density, characteristic of the solar core.
We compare these results with those obtained by BL and MP.
In \S5, we give the results for the heavy element diffusion coefficients,
obtained under the same conditions.
In \S6, we give the diffusion velocities in the sun, and
again we compare our results with those obtained by BL and MP.
In \S7, we give analytical
expressions for our numerical results. In \S8, we compare our expression
for the electric field with the value obtained by Braginskii.
Finally, in \S9, we give a summary of the most important results.

\sect{2. Basic Equations}

Each species of particles $s$ is described by a distribution function
$F_s({\bf x},{\bf v},t)$ normalized to unit integral, a mean number
density $n_s$, an ionic charge $q_s\equiv Z_se$, and a mass $m_s$. All
species are assumed to be at the same temperature $T$ and in an
overall hydrostatic equilibrium, since the temperature and pressure
equilibration timescales are much shorter than the diffusion times.
The mass and
charge densities are $\rho_s=n_sm_s$ and $\rho_{es}=n_sq_s$.
The mean fluid
velocity of each species is defined by
$${\bf u}_s=\int{\bf v}F_sd{\bf v}.\eqno(\e)$$
The mean fluid velocity is given by
$${\bf u}={1\over\rho}\sum_s\rho_s{\bf u}_s,\eqno(\e)$$
where $\rho=\sum_s\rho_s$ is the total mass
density. The diffusion velocity for species $s$ is defined by
$${\bf w}_s={\bf u}_s-{\bf u},\eqno(\e)$$
and is therefore measured relative to the mean velocity
of the fluid as a whole.
We define the``residual heat flow vector'' for species
$s$ by (Burgers, 1969):
$${\bf r}_s=\biggl[{m_s\over 2k_BT}\int F_s({\bf v}-{\bf u})|{\bf v}-
       {\bf u}|^2d{\bf v}-{5\over
2}{\bf w}_s\biggr],\eqno(\e)$$
where $k_B$ is Boltzmann's
constant. The cross-section for Coulomb scattering between particles of
species $s$ and
of species $t$ ($s$ can be equal to $t$) is given by
$$\sigma_{st}=2\sqrt{\pi}e^4Z_s^2Z_t^2(k_BT)^{-2}{\ln}\Lambda_{st},\eqno(\e)$$
where ${\ln}\Lambda_{st}$ is
the Coulomb logarithm ($\Lambda_{st}$ is the ``plasma parameter''),
a correction factor taking into account
the logarithmic contribution of binary encounters with impact parameters up to
the Debye shielding length.
For the
Coulomb logarithm, we adopt the following expression, obtained by
Iben and MacDonald (1985) using numerical results from Fontaine and Michaud
(197
 9b),
$${\ln}\Lambda_{st}={1.6249\over 2}{\ln}\left[1+0.18769\left(
{4k_BT\lambda\over Z_sZ_te^2}\right)\right],\eqno(\en\coullogii)$$
where $\lambda={\rm max}(\lambda_D,a_0)$,
$\lambda_D=(k_BT/4\pi e^2\sum_s n_sZ_s^2)^{1/2}$
is the Debye length, and $a_0=(3/4\pi\sum_{ions}n_i)^{1/3}$ is the interionic
di
 stance.
The friction coefficient between species $s$ and $t$ is
$$K_{st}=(2/3)\mu_{st}(2k_BT/\mu_{st})^{1/2}\,n_sn_t\sigma_{st}\eqno(\en\Kst)$$
where $\mu_{st}\equiv m_sm_t/(m_s+m_t)$ is the reduced   mass for species $s$
and $t$.

The Burgers equations for mass, momentum, and energy conservation can then
be written as
$${\partial n_s\over\partial t}+{1\over
         r^2}{\partial\over\partial r}(r^2 n_s w_s)=
         \biggl ({\partial n_s\over\partial t}\biggr)_{nucl.},\eqno(\en\diff)$$
$${dp_s\over
dr}+\rho_sg-\rho_{es}E=\sum_{t\ne s}K_{st}
     \Bigl[(w_t-w_s)+0.6(x_{st}r_s-y_{st}r_t)\Bigr],
           \eqno(\en\burgerspressure)$$
and
$${5\over 2}n_sk_B{dT\over dr}=\sum_{t\ne s}K_{st}\bigl\{{3\over 2}x_{st}
           (w_s-w_t)
    -y_{st}\Bigl[1.6x_{st}(r_s+r_t)+Y_{st}r_s-4.3x_{st}r_t\Bigr]\Bigr\}
    -0.8K_{ss}r_s.\eqno(\en\burgersheat)$$
In these equations, ${\bf g}\equiv -(GM(r)/r^2)\hat e_r$ is the gravitational
acceleration,
$g\equiv |{\bf g}|$, ${\bf E}$ is the
electric field, $E\equiv |{\bf E}|$, $x_{st}=\mu_{st}/m_s$,
$y_{st}=\mu_{st}/m_t$,
and
$Y_{st}=3y_{st}+1.3x_{st}m_t/m_s$.
The numerical coefficients in equations~(\burgerspressure) and~(\burgersheat)
ar
 e
related to the collision integrals and were obtained using a pure Coulomb
potential with a long-range cutoff at the Debye length. More accurate results
can be obtained by using numerical values derived from calculations using a
scre
 ened
Coulomb potential.
We have assumed spherical symmetry and
included a term for composition changes due to
nuclear burning in equation~(\diff).
Using equation~(\burgerspressure), it is straightforward to show that
$$\sum_s\biggl({dp_s\over dr}+\rho_sg-\rho_{es}E\biggr)=0,\eqno(\e)$$
or
$${dp\over dr}+\rho g-\rho_e E=0\eqno(\en\momentumcons)$$
where $p\equiv\sum_sp_s$ and  $\rho_e\equiv\sum_s\rho_{es}$ are the
total pressure and total charge density.
The departure from local charge neutrality is
very small, with $\rho_e E/\rho g\sim Gm_p^2/e^2\sim 10^{-37}$ (see discussion
and eqs.(22)-(23) in BL).
Equation~(\momentumcons) therefore reduces to the familiar equation of
hydrostatic equilibrium,
$$dp/dr=-\rho g.\eqno(\en\hydroequ)$$ In addition, the following
constraints must be satisfied:
charge neutrality,
$$\sum_sq_{es}n_s=0;\eqno(\en\charge)$$
current neutrality,
 $$\sum_sq_{es}n_sw_s=0,\eqno(\en\current)$$ and
local mass conservation,
$$\sum_sm_sn_sw_s=0.\eqno(\en\velocity)$$
Note that equation~(\velocity) follows from the fact that the diffusion
equations are
solved in the rest frame of the plasma.
The set of linear equations (\burgerspressure)-(\burgersheat)
and (\current)-(\velocity) forms a closed system for the diffusion velocities
${\bf w}_s$, the residual heat flow vectors ${\bf r}_s$, the gravitational
acceleration ${\bf g}$ and
the electric field ${\bf E}$ in terms of the pressure,
temperature, and concentration gradients. Since we already know the value of
$g=GM(r)/r^2$, this relation provides a useful check on the numerical results.

If we ignore thermal diffusion, the electric field is given by
$E=-(1/en_e)(\partial p_e/\partial r)$. When thermal effects are
included, the electric field
can be written as (see eq.~(57) in BL)
$$eE=-{1\over n_e}\,{\partial p_e\over \partial r}-\alpha_ek_B
{\partial T\over\partial
r}.\eqno(\en\Efield)$$
Using his two-component fluid equations, Braginskii (1965) has obtained
the values $\alpha_e\approx 0.71$ for
a pure hydrogen plasma and $\alpha_e\approx 0.9$ for a pure helium plasma.

\sect{3. Method of solution}

The system of equations (\burgerspressure)-(\burgersheat) and
(\current)-(\velocity) can be solved numerically.
If there are $S$ species in the system ($S-1$ ions plus electrons), there
are $S$ momentum equations~(\burgerspressure) and $S$
energy equations~(\burgersheat). The unknowns
are the $S$ drift velocities $w_s$ and the $S$ heat
fluxes $r_s$. The gravitational acceleration and the electric field are
also treated as unknowns, and we use
the two additional equations for mass and charge conservation,
equations~(\current) and (\velocity),
to help determine $g$ and $E$.
Note that the hydrostatic equilibrium condition (eq.~\hydroequ) should
be satisfied automatically, providing a useful check on the numerical results.

We now rewrite the basic equations in a dimensionless form that is
better suited to a numerical treatment.
The radius $r$ is expressed in units of $R_\sun$, the mass density $\rho$ in
units of
$100\,{\rm g cm}^{-3}$ and temperature $T$ in units of $10^7\,{\rm K}$,
characteristic values at the
center of the sun, and the time $t$ is in units of $\tau_0=6\times 10^{13}
{\rm yrs}$, a
characteristic diffusion time in the sun (see,e.g., Kippenhahn and Weigert
p.60, or eq.~(9) in BL).
We write the Burgers equations (\burgerspressure)-(\burgersheat) and the
constraints
(\current)-(\velocity) as
$${p\over K_0}\Bigl[\alpha_i{d\ln p\over dr}+\nu_i{d\ln T\over dr}
      +\sum_{\scriptstyle{j=1}\atop\scriptstyle{j\ne e,2}}^S\gamma_{ij}
{d\ln C_j\over dr}\Bigr]
       =\sum_{j=1}^{2S+2}\Delta_{ij}W_j,\eqno(\en\system)$$
where the following notations have been introduced.
The concentration of species $s$ is defined by
$$C_s\equiv n_s/n_e.\eqno(\en\cs)$$
It is related to the mass fractions $X_s\equiv m_sn_s/\rho$
by
$$C_s={X_s/A_s\over\sum_iZ_iX_i/A_i}\eqno(\e)$$
or inversely
$$X_s={A_sC_s\over\sum_iA_iC_i},\eqno(\e)$$
where $A_s$ is the atomic number of species $s$, and the sum is over all
species,
including the electrons. For the electrons,
$$A_e\equiv m_e/m_0,\eqno(\en\Ae)$$
where $m_e$ and $m_0$ are
the electron and atomic masses, and
$$C_e\equiv 1.\eqno(\en\ce)$$
The constant $K_0$ is given by
$$K_0=1.144\times 10^{-40}\,T^{-3/2} n_e^2,\eqno(\e)$$
where $T$ and $\rho$ are expressed in the characteristic units defined above.
We use the ideal gas equation of state, $p_s=n_sk_BT$, and
equations~(\cs),(\Ae) and (\ce) to write
$${p\over K_0}=2.00\,{T^{5/2}\over\rho}\,(\sum_sC_s)(\sum_sA_sC_s),
\eqno(\en\pK)$$
where we have  written
the electron number density in terms of the mass density,
$\rho=m_0n_e\sum_sA_sC_s$.
The variables $W_i$ are
$$W_i=\cases{w_i & for $i=1,...S$\cr
             r_{i-S} & for $i=S+1,...2S$\cr
             K_0^{-1}n_eeE & for $i=2S+1$\cr
             K_0^{-1}n_em_0g & for $i=2S+2$.\cr}\eqno(\e)$$
If we define $C\equiv\sum_iC_i$,
the coefficients on the left-hand-side of equation~(\system) are given by
$$\alpha_i=\cases{\displaystyle{C_i\over C} &for $i=1,2,...S$ \cr
                  0  &for $i=S+1,...2S+2$,\cr}\eqno(\e)$$
$$\nu_i=\cases{2.5\,\displaystyle{C_{i-S}\over C} &for
                          $i=S+1,...2S$\cr
                0 &for $i=1,...,S$ and $i=2S+1,2S+2$,\cr}\eqno(\e)$$
and
$$\gamma_{ij}=\cases{\displaystyle{C_i\over
          C}\biggl[\biggl(\delta_{ij}-\displaystyle{C_j\over C}\biggr)-
            \biggl(\delta_{i2}-\displaystyle{C_2\over C}\biggr)
          \displaystyle{Z_jC_j\over Z_2C_2}\biggr]
                           &for $i=1,...,S$\cr
                      0 &for $i=S+1,...,2S+2$.\cr}\eqno(\e)$$
The coefficients on the right-hand-side of equation~(\system) are given by
$$\Delta_{ij}=\cases{-\sum_{k\ne i}\kappa_{ik} &for $j=i$\cr
                     \kappa_{ij} &for $j=1,...,S$ and $j\ne i$\cr
                     \sum_{k\ne i}0.6\kappa_{ik}x_{ik} &for $j=i+S$\cr
                     -0.6\kappa_{i,j-S}y_{i,j-S} &for $j=S+1,...,2S$
                               and $j\ne i+S$\cr
                     Z_iC_i &for $j=2S+1$\cr
                     -A_iC_i &for $j=2S+2$\cr}\eqno(\e)$$
for i=1,...,S, by
$$\Delta_{ij}=\cases{\sum_{k\ne j}1.5\kappa_{i-S,k}x_{i-S,k} &for $j=i-S$\cr
                     -1.5\kappa_{i-S,j}x_{i-S,j} &for $j=1,...,S$
                             and $j\ne i-S$\cr
                     -\sum_{k\ne i}\kappa_{i-S,k}y_{i-S,k}
                             (1.6x_{i-S,k}+Y_{i-S,k})\cr
                       \hskip1cm -0.8\kappa_{i-S,i-S} &for $j=i$\cr
                     2.7\kappa_{i-S,j-S}y_{i-S,j-S}x_{i-S,j-S}
                             &for $j=S+1,...,2S$ and $j\ne i$\cr
                     0 &for $j=2S+1,2S+2$\cr}\eqno(\e)$$
for i=S+1,...2S, by
$$\Delta_{ij}=\cases{Z_jC_j &for $j=1,...,S$\cr
                      0 &for $j=S+1,...,2S+2$\cr}\eqno(\e)$$
for i=2S+1, and finally by
$$\Delta_{ij}=\cases{A_jC_j &for $j=1,...,S$\cr
                      0 &for $j=S+1,...,2S+2$\cr}\eqno(\e)$$
for i=2S+2.
In these expressions, the coefficient $\kappa_{st}$ is defined by
$$\kappa_{st}=\biggl({A_sA_t\over
     A_s+A_t}\biggr)^{1/2}C_sC_tZ_s^2Z_t^2{\ln}\Lambda_{st}.
\eqno(\en\kappast)$$
It is related to the friction coefficient through $K_{st}=K_0\kappa_{st}$.
We have used
the constraint of charge
neutrality to eliminate the
concentration gradient of species 2 in equation~(\system),
$${d{\ln}C_2\over dr}=-\sum_{\scriptstyle{j=1}\atop\scriptstyle{j\ne e,2}}^S
                         {Z_jC_j\over Z_2C_2}\,{d{\ln}C_j\over dr}.
        \eqno(\en\gradient)$$

Since $p/K_0$ is proportional to $T^{5/2}/\rho$ (see eq.~\pK),
all the velocities will be
proportional to $T^{5/2}/\rho$. Therefore, we  introduce the function
$\xi_s$ (following BL), such
that $$w_s=(T^{5/2}/\rho)\xi_s.\eqno(\e)$$
The rate of change of the element mass fractions due to diffusion is now
written in
dimensionless form as
$${\partial X_s\over\partial t}=-\,{1\over \rho
           r^2}{\partial\over \partial r}[r^2X_sT^{5/2}\xi_s(r)],
            \eqno(\en\xieq)$$
the generalization of equation~(1) of BL to the case of arbitrary
concentrations and a more accurate treatment of the plasma physics.

Equations~(\system) are linear. Therefore, we can combine linearly the
solutions obtained by
keeping only one of the
gradients different from zero.
We write the results in terms of
generalized diffusion coefficients
$A_p(s)$, $A_T(s)$ and
$A_t(s)$ for species $s$, as
$$\xi_s(r)=A_p(s){\partial \ln p\over \partial r}+A_T(s){\partial \ln T\over
         \partial r}+\sum_{t\ne e,2} A_t(s){\partial\ln C_t\over \partial r}.
\eqno(\en\xis)$$
If ${\ln}\Lambda$ is
assumed identical for all the interactions, the coefficients $A_p$, $A_T$,
and $A_t$ are
functions of the mass fractions only. If ${\ln}\Lambda$ is defined by
equation~(\coullogii), these coefficients also depend
on the charges, the temperature, and
the density.

\sect{4. Hydrogen and helium diffusion}

First, we consider the diffusion of hydrogen and helium, neglecting
the presence of heavier elements.
We calculate the hydrogen diffusion velocity. The helium
diffusion velocity is then simply obtained from the constraint that there is no
mean fluid velocity, $\sum_sX_sw_s=0$. Neglecting the electron mass compared to
the proton mass, we have
$w_\alpha=-(X/Y)w_H$. The rate of change of the helium number density is
therefore equal and opposite to the rate of change of the hydrogen
number density, $(\partial Y/\partial t)=-(\partial X/\partial t)$. Helium
diffuses towards the center of the star, whereas hydrogen diffuses outwards.

In the absence of heavy elements, the function $\xi_H$ is given
by\baselineskip=12pt\footnote{$^3$}{Note
that BL define $\xi_H$ with the opposite sign. They also write $\xi_H$ in
terms of the mass
fraction gradient instead of the number concentration gradients. These are
simply related by
$\partial\ln C_s/\partial r=\partial\ln X_s/\partial r-(\sum_iZ_iX_i/A_i)^{-1}
\sum_j(Z_jX_j/A_j)
\partial\ln X_j/\partial r$
or inversely,
$\partial\ln X_s/\partial r=\partial\ln C_s/\partial r-(\sum_iA_iC_i)^{-1}
\sum_jA_jC_j\,\partial\ln C_j/\partial r$.
If hydrogen and helium are the only elements,
$\partial\ln X/\partial r=(1+X)\partial\ln C_H/\partial r$.}\baselineskip=18pt
$$\xi_H=A_p(H){\partial \ln p\over\partial r}+
 A_T(H){\partial\ln T\over\partial r}+A_H(H){\partial\ln C_H\over\partial r}.
\eqno(\en\xiH)$$
We have chosen
helium as element number 2, i.e., we write the
diffusion velocity in terms of the hydrogen concentration
gradient, using equation~(\gradient) to eliminate the
helium concentration gradient.

Two major simplications are usually made when calculating the
hydrogen and helium diffusion velocities in the absence of heavy elements.
It is usually assumed that the Coulomb logarithms ${\ln}\Lambda_{ij}$ are
 identical for
all interactions. This allows the factorization
of $\ln\Lambda$ outside the function $\xi_H$
(see, eg., Noerdlinger 1977, Bahcall and Loeb 1990).
In that case, the coefficients $A_p$, $A_T$ and $A_H$ depend only on the
hydrogen (or helium) concentration, not on density, temperature and ionic
charges.
The second simplification is to ignore the residual heat fluxes ${\bf r}_s$.
Then, the diffusion velocities are easier to calculate analytically, since
there is no need for
the heat equations~(\burgersheat) and the
number of variables and equations is reduced by a factor of two.
However, these simplifications can lead to large relative errors in the
diffusion
velocities. In particular
it has been argued by MP that thermal diffusion can increase the diffusion
velocities
by 30\%.

In figure~1a-c, we show the variation of the coefficients $A_p$, $A_T$,
and $A_H$ with the hydrogen mass fraction $X$.
To obtain these
results, we have assumed $T=10^7\,{\rm K}$ and
$\rho=100\,{\rm g~cm}^{-3}$,
typical values in the core of the sun.
The exact results are
represented with solid lines,
the results obtained neglecting the heat fluxes are
represented with
short-dashed lines, and the results obtained by keeping the heat fluxes but
usin
 g
$\ln\Lambda=2.2$ for all interactions\baselineskip=12pt\footnote{$^4$}{This
value is usually
considered representative of the Coulomb logarithms
in the solar interior (see, e.g., Noerdlinger 1977 and
BL).}\baselineskip=18pt
are represented with
long-dashed lines.
If the heat fluxes are totally neglected,
the two coefficients $A_p$ and $A_H$ are underestimated, and
$A_T=0$.
In figure~1d, we show the relative errors on the coefficients due to
these approximations. The short-dashed line represents the error
on $A_p$ and $A_H$ when the heat fluxes are neglected.
It can be as high as 45\% for small values of the hydrogen
mass fraction (not relevant to the sun).
The long-dashed lines represent the errors
due to $\ln\Lambda=2.2$.
The errors on $A_p$, $A_H$ and $A_T$ are
smaller than  20\%, except when $X\sim 1$.
In the  interior of the sun, $X$ varies approximately
between 0.3 and 0.7.
For these values of $X$, the error does not exceed 20\%.

\subsect{4.1 Comparison with Bahcall and Loeb}

In order to keep the analytical calculations simple without neglecting the
thermal effects one can use
``effective''
thermal diffusion
coefficients (obtained through fits to the exact numerical results).
In figure~2a, we show the ratio between the ``exact''  coefficients and those
obtained
by BL who neglected the residual heat fluxes, assumed a Coulomb logarithm of
2.2
  for
all the interactions, and used an ``effective'' thermal diffusion coefficient.
The expressions obtained by BL are:
$$\xi_H^{BL}(r)=A_p^{BL}{\partial\ln p\over\partial r}+A_T^{BL}{\partial\ln
T\over\partial r}+A_H^{BL}{\partial\ln C_H\over\partial r}\eqno(\en\xibl)$$
with
$$A_p^{BL}=-5(1-X)/4,\eqno(\en\apbl)$$
$$A_T^{BL}=-6(1-X)(X+0.32)/(1.8-0.9X)(3+5X),\eqno(\en\atbl)$$
and
$$A_H^{BL}=-(X+3)/(3+5X).\eqno(\en\ahbl)$$
The result for the thermal diffusion coefficient was obtained by fitting values
obtained previously by Aller and Chapman (1960), Montmerle
and Michaud (1976), and Noerdlinger (1978).
For values of $X$ between 0.3 and 0.7, the error made by BL on $A_p$
and $A_H$ is smaller than 40\%, whereas the error on $A_T$ is as large
as 70\%. However, as we will show in \S 6, large errors on $A_T$ do not
necessarily
lead to large errors on the diffusion velocities, since the temperature
gradient in the sun
is much smaller than the pressure and concentration gradients.

It is important to notice that the heat fluxes
affect not only the thermal diffusion coefficients,
but also the pressure
and composition gradient coefficients.

\subsect{4.2. Comparison with Michaud and Proffitt}

MP solved the Burgers equations with and without including the heat fluxes,
then represented
the effects of the heat fluxes
by an ad-hoc correction to the results obtained
when neglecting those heat fluxes. In our dimensionless variables, their
results can be written as
$$\xi_H^{MP}(r)=A_p^{MP}{\partial\ln p\over\partial r}+A_T^{MP}{\partial\ln
T\over\partial r}+A_H^{MP}{\partial\ln C_H\over\partial r}\eqno(\en\ximp)$$
with
$$A_p^{MP}=-{5\over 4}{(1-X)\over (0.7+0.3X)(\ln\Lambda_{xy}/2.2)},
\eqno(\en\apmp)$$
$$A_T^{MP}=-{9\over 8}{(1-X)\over (0.7+0.3X)(\ln\Lambda_{xy}/2.2)},
\eqno(\en\atmp)$$
and
$$A_H^{MP}=-{(X+3)\over
(3+5X)(0.7+0.3X)(\ln\Lambda_{xy}/2.2)},\eqno(\en\ahmp)$$
where
$$\ln\Lambda_{xy}=-19.95-{1\over 2}\ln\rho-{1\over 2}\ln{X+3\over 2}+
{3\over 2}\ln T.\eqno(\en\lnmp)$$

In figure~2b, we show in solid lines
the ratio of our coefficients and those obtained by
MP.
The difference between our results and those obtained by MP is smaller than
15\%
  for
$A_T$, and smaller than 5\% for $A_p$ and $A_H$.
This small discrepency will be discussed
in \S6.

\sect{5. Heavy element diffusion}

Because of the complexity caused by the addition of heavy elements,
this problem has always been
approached with additional simplifications (Vauclair {\it et al.} 1974,
Noerdlinger 1978, BL, MP).
One common simplification is to
assume a negligible
helium concentration, therefore reducing the problem to a
two-species situation.
However, this assumption is not valid
in the interior of the sun, where the characteristic mass fractions of
hydrogen,
helium, and oxygen are of the order of
0.34, 0.64 and 0.01  respectively (see, e.g., Bahcall 1990).

The functions $\xi_s$ for the hydrogen and oxygen are now written as
$$\xi_H=A_p(H){d\ln p\over dr}+A_T(H){d\ln T\over T}+A_H(H){d\ln C_H\over dr}+
A_O(H){d\ln C_O\over dr},\eqno(\en\xiHH)$$
and\baselineskip=12pt\footnote{$^5$}{Note
that BL define the function $\xi_A$ such that
$\partial Z/\partial t=-(1/\rho r^2)\,\partial[r^2XZT^{5/2}\xi_A/(2-X)]
\partial r.$
The function $\xi_A$ is related to the function $\xi_O$ through
$\xi_A=\xi_O (2-X)/X$.}\baselineskip=18pt
$$\xi_O=A_p(O){\partial \ln p\over \partial r}+A_T(O){\partial \ln T\over
         \partial r}+A_H(O){\partial\ln C_H\over\partial r}
          +A_O(O){\partial\ln C_O\over \partial r}.\eqno(\en\xiOO)$$

It is interesting to show the variation of the diffusion coefficients as a
function of
the hydrogen mass fraction for fixed values of the temperature, density, and
heavy element
mass fraction. Indeed, as we will show in the next section, in the sun
 these parameters
all vary with radius, and it is
more difficult to extract the hydrogen mass fraction dependence itself.
In figure~3, we show the four coefficients $A_p(H)$, $A_T(H)$, $A_H(H)$ and
$A_O(H)$
as a function of the oxygen mass
fraction $X$.
These results were obtained using $T=10^7\,{\rm K}$, $\rho=100\,{\rm
g~cm^{-3}}$
and $Z=0.01$, where $Z$ is the oxygen mass fraction. These values are typical
in the solar interior.
Note that now $X$ has a maximum value determined by $X_{max}=1-Z$, because
of the charge
neutrality constraint. We notice that
the coefficient $A_O(H)$ is two orders of magnitude smaller than the other
coefficients. This was expected, since we
have chosen $K_{HO}/K_{H\alpha}\sim Z/Y\sim 10^{-2}$.
The error made by neglecting the presence of oxygen
when calculating the hydrogen diffusion velocity
is smaller than 2\%.

In figure~4 we show the oxygen diffusion
coefficients $A_p(O)$, $A_T(O)$, $A_H(O)$ and $A_O(O)$.
Again, the coefficient $A_O(O)$ is much smaller than the other three
coefficients,
and can be neglected to the level of precision desired in these calculations.

\sect{6. Diffusion velocities in the sun}

We now calculate the diffusion velocities of hydrogen, helium, and heavier
elements
in the present solar interior ($r<0.7R_\sun$). We use values for the pressure,
temperature,
density and mass fractions of the contemporary sun,  obtained from
the standard solar model (table 4.4, in Bahcall 1990).
Since we have the radial
profile of the pressure, temperature, and mass fractions, we can calculate
their gradients.
The coefficients $A_p$, $A_T$, $A_H$ and $A_O$ are computed using
the tabulated values of $T$, $\rho$, $X$, $Y$, and $Z$.
The iron abundance is assumed to be uniform and given by ${\rm log}_{10}
(n_{Fe}/n_H)=6.82-12$,
and its ionization is 21.

In figure~5, we show the radial variation of the
hydrogen diffusion coefficients.
In figure~6, we show the relative importance of the different terms in the
hydrogen diffusion velocity.
The oxygen concentration gradient gives a negligible contribution to
$w_H$. The temperature term is not negligible, but is smaller than the
pressure term
(between 25\% and 50\% of the pressure term).
Therefore, a large error on the temperature diffusion coefficient does not
necessarily lead to a large error for the diffusion velocities.

The timescale for a change in the element abundances can be
characterized by $t\equiv r/w_H$.
To obtain the time $t$ in units of $t_0=5\times 10^9\,{\rm yrs}$,
the age of the sun, we simply multiply the dimensionless time by $t_0/\tau_0$,
where
$\tau_0$ is the characteristic diffusion time defined in \S~3.
The smaller the time $t$, the faster the element concentrations change.
In figure~7, we show the variation of $t$ with the radius.
The fastest changes in the hydrogen concentration occurs at approximately
$0.05R_\sun$, where
$t_{min}\sim 70\,t_0$.

\subsect{6.1. Comparison with Bahcall and Loeb}

In figure~8a, we compare our exact results with those obtained
using the analytic BL formulae, equations~(\apbl)-(\ahbl) (eqs. 1-5 in BL).
The BL formula underestimate the diffusion coefficients.
The error
on the pressure and
concentration diffusion coefficients
is smaller than $30\%$. The error on the
temperature diffusion coefficient is of the order of $50\%$, except near the
center where it becomes very large.

In figure~9a-c, we show in solid lines the results for the diffusion
velocities of
hydrogen, oxygen, and iron.
The helium diffusion velocity $w_\alpha$ is related to the hydrogen and
oxygen diffusion
velocities through the zero mean velocity constraint, equation~(\velocity),
$$w_\alpha=-(X_Hw_H+X_Ow_O+X_{Fe}w_{Fe})/X_{He}.\eqno(\e)$$
We also show in short-dashed lines the BL results for the hydrogen and
oxygen diffusion velocities, given by
equations~(\apbl)-(\ahbl) for the hydrogen velocity, and equations~(2)-(5)
in BL for the
oxygen velocity.
In the BL approximation, the helium velocity is given by
$w_\alpha=-(X_H/X_{He})w_H$.
The error
in $w_H$ due to the BL approximations
is smaller than 30\%, except near the center,
where
the error is as high as 70\%. However, the
error for the oxygen diffusion velocity is very large. This was expected,
since BL neglected
completely the presence of helium when
calculating $w_O$. The BL results for $w_O$ are therefore only valid when
$X\approx 1$.

\subsect{6.2. Comparison with Michaud and Proffitt}

In figure~8b, we compare our results with those obtained using
the MP formula, equations~(\apmp)-(\ahmp).
The difference between our results and the results obtained by MP for the
diffusion coefficients is smaller than 5\% for $A_p$ and $A_H$. Our thermal
diffusion coefficient $A_T$ is larger by about 20\%. This discrepency may
result from the fact that we have used $z_{st}=0.6$ for the heat flux
coefficien
 t
in equations (\burgerspressure) and (\burgersheat).
If the collision
integrals are calculated using a screened Coulomb potential, the value obtained
for $z_{st}$ is about $2/3$ smaller (Proffitt 1993).

In figure~9a-c, we show the results for the diffusion velocities
of hydrogen, oxygen and iron. The MP result for the heavy elements
diffusion velocities
are given by
$$\eqalign{\xi_i=-{2\over\sqrt{5}Z_i^2}&
\left[
{d\over dr}\left\lbrace\ln\left[{X_i\over 5X+3}
\left({1+X\over 5X+3}\right)^{Z_i}\right]
\right\rbrace
+\left[1+Z_i-A_i\left({5X+3\over 4}\right)\right]{d\ln p\over dr}
\over
X(A_{ix}^{1/2}C_{ix} - A_{iy}^{1/2}C_{iy} ) + A_{iy}^{1/2}C_{iy}
\right]\cr
&+X\xi_H\left[{(A_{ix}^{1/2}C_{ix} - A_{iy}^{1/2}C_{iy} ) \over
X(A_{ix}^{1/2}C_{ix} - A_{iy}^{1/2}C_{iy} ) + A_{iy}^{1/2}C_{iy} }
-0.23\right]\cr
&+{0.54(4.75X+2.25)\over(\ln\Lambda_{xy}+5)}{d\ln T\over dr},\cr}\eqno(\e)$$
where $A_{ij}\equiv A_iA_j/(A_i+A_j)$ is the reduced mass in
atomic number units,\hfill\break
$C_{ij}= \ln[\exp(1.2 \ln\Lambda_{ij}) +1] /1.2$, and
$\ln\Lambda_{ij}$ is given by equation~(\lnmp).
The difference between our results and those obtained by MP is smaller than
15\%
  for the
hydrogen diffusion velocity, and smaller than 20\% for the oxygen
diffusion velocity.

\sect{7. Analytical fits}

All the results shown above were obtained using
the expression~(\coullogii) for the ``effective'' Coulomb
logarithms.  If we use (as in the previous results)
the correct charge, temperature and density dependent expression for the
Coulomb logarithms, we cannot give a simple analytical fit, as
in BL, for the diffusion coefficients in terms of
the hydrogen mass fraction. Even though one could in principle incorporate
a subroutine which  solves the problem of element diffusion in a
standard solar model evolution code,
it is useful to give a simple analytical fit of the results obtained
here in the solar interior.
It is convenient to provide to stellar evolution programs
other types of input physics, such as opacities or equations
of state, in the form of tabulated values, or in terms of approximate
analytical fits.
In order to provide a similarly convenient service for diffusion,
we adopted a constant for each  Coulomb logarithm,
the value it has at the center of the sun.
In figure~10, we show the values of the various Coulomb logarithms
in the sun computed
using equation~(\coullogii).
In figure~11a-c, we compare
the diffusion coefficients  obtained with these
Coulomb logarithms with those obtained
using a constant value of $\Lambda$ (equal to its central value).
In figure~11d, we show that the error on $w_H$ is smaller than
4\% in the solar core ($r/R_\sun\le 0.4$), and remains smaller than 15\%
up to the convection zone. The error on the heavy elements diffusion
velocities are $\simless 6\%$ in the solar core, and remain $\simless 20\%$ up
t
 o
the convection zone.
We can now fit these results to second order
polynomials. Since the presence of oxygen and iron
have very little influence on the hydrogen velocity,
we can assume that the diffusion coefficients depend only on $X$. We find:
$$\left\{\eqalign{&A_p(H)=-2.09+3.15\,X-1.07\,X^2,\cr
        &A_T(H)=-2.18+3.12\,X-0.96\,X^2,\cr
        &A_H(H)=-1.51+1.85\,X-0.85\,X^2,\cr}\right.\eqno(\en\fith)$$
for the hydrogen diffusion coefficients,
$$\left\{\eqalign{&A_p(O)=0.15+1.34\,X-0.89\,X^2,\cr
        &A_T(O)=0.53+1.99\,X-0.72\,X^2,\cr
        &A_H(O)=0.08+0.58\,X-0.28\,X^2,\cr}\right.\eqno(\en\fito)$$
for the oxygen diffusion coefficients, and
$$\left\{\eqalign{&A_p(Fe)=0.25+1.31\,X-0.87\,X^2,\cr
        &A_T(Fe)=0.65+1.99\,X-0.75\,X^2,\cr
        &A_H(Fe)=0.09+0.53\,X-0.27\,X^2,\cr}\right.\eqno(\en\fitf)$$
for the iron diffusion coefficients.
The errors due to the polynomial fits are of the order of 0.2\%.
There is no
need to use higher order polynomials, since the error made with these second
order polynomials are already much smaller
than the errors introduced by the simplified Coulomb logarithm.

For an easy comparison with BL and MP results, we can factorize the BL
results and write
the numerical fits as
$$A_p(H)=-{5\over 4}(1-X)a_p,\eqno(\en\bli)$$
$$A_T(H)=-{6(1-X)(X+0.32)\over (1.8-0.9X)(3+5X)}a_t,\eqno(\e)$$
and
$$A_H(H)=-{(X+3)\over(3+5X)}a_H.\eqno(\e)$$
First order polynomial fits give the following  analytical results:
$$\left\{\eqalign{&a_p=1.66-0.82\,X,\cr
   &a_T=4.46-3.65\,X,\cr
   &a_H=1.63-0.74\,X.\cr}\right.\eqno(\en\blf)$$
The error introduced by the polynomial fit is much smaller than the error due
to the simplified Coulomb logarithm.
Equations~(\bli)-(\blf) can be used to improve existing diffusion subroutines
that are based on the BL formalism.

It is important to remember that these fits are made for the standard
parameters of
the solar interior.

\sect{8. Hydrostatic equilibrium and electric field}

As explained in \S~2, we solve the system of equations~(\system) for the
diffusion velocities, the residual heat flow vectors, the gravitational
acceleration, and the electric field.
On the other hand, the equation of hydrostatic equilibrium
given by equation~(\hydroequ) must be satisfied.
The numerical results for $g$ can be written as
$$g=-{p\over\rho}\biggl[A_p(g){d\ln p\over dr}+A_T(g){d\ln T\over dr}
+A_H(g){d\ln C_H\over
dr}\biggr].\eqno(\e)$$
In hydrostatic equilibrium, $A_T=A_H=0$ and $A_p=1$.
The equation of hydrostatic equilibrium is satisfied
to $10^{-7}$ by the numerical results.

Similarly, we write the results for the electric field as
$$E={-p_e\over en_e}\biggl[A_p(E){d\ln p\over dr}+A_T(E){d\ln T\over dr}+
A_H(E){d\ln C_H\over
dr}\biggr].\eqno(\en\electricfield)$$
The coefficients in equation~(\electricfield) should be compared with
equation~(\Efield).
We find an excellent agreement between our numerical results and
equation~(\Efield)
for the coefficients in front of
the pressure and concentration gradients (with an error smaller than
$0.2\times 10^{-2}$).
Our values for the coefficient $A_T(E)$ are slightly
larger than the results obtained by Braginskii (1965) for $\alpha_e$. As
shown in
figure~12, we observe however
the same tendency for this coefficient to increase slightly with an increase
in the helium
concentration. We get $\alpha_e=0.8$ for $X=1$, and $\alpha_e=0.9$ for $X=0$.

\sect{9. Summary and conclusions}

We have developed a Fortran program to solve numerically
the Burgers equations for an
arbitrary number of species, without any approximation.
For the discussion of solar conditions given here,
we have neglected the radiative forces,
but these forces could easily be incorporated in the numerical routine.
The accuracy of the results for the diffusion velocities is limited
only by the validity of the expression for the Coulomb logarithm.
The diffusion velocities of hydrogen, oxygen and iron were calculated in
the solar interior and compared with the results of
Bahcall and Loeb (1990) and by Michaud and Proffitt (1992).
The results of BL for the hydrogen diffusion velocity are smaller by
$\sim 30\%$, except near the center, where the error is much larger.
The results obtained by MP for the hydrogen and oxygen diffusion velocities
differ by $\simless 15\%$.

We provide analytical fits of our numerical results for the diffusion
coefficients
as a function of the hydrogen mass fraction  in the solar interior (eqs.
\fith-\
 blf).
These fits were obtained by assuming fixed values for the Coulomb
logarithms, equal to their values at the center of the sun.

\ack{
We are grateful to M. H. Pinsonneault and C. R. Proffitt for valuable
discussions and suggestions.
This work was supported by NSF Grant PHY92-45317.}

\vfill\eject

\Refs
\ref\apj{Aller, L. H., and Chapman, S. 1960}{132}{461}
\refbook Bahcall, J. N. 1990, {\it Neutrino Astrophysics} (Cambridge
            University Press, Cambridge).
\ref\apj{Bahcall, J. N., and Loeb, A. 1990}{360}{267}
\ref\apj{Bahcall, J. N. and Pinsonneault, M. H. 1992a}{395}{L119}
\ref\rmp{\bysame 1992b}{64}{885}
\refbook Book, D. L. 1990, {\it NRL Plasma Formulary} (NRL Publication
              177-4405).
\refbook Braginskii, S. I. 1965, {\it Reviews of Plasma Physics}, Vol.1, p.205
               (Consultants
                     Bureau, New York).
\refbook Burgers, J. M. 1969, {\it Flow Equations for Composite
               Gases} (Academic Press, New York).
\ref\apj{Chaboyer, B., Deliyannis, C. P., Demarque, P., Pinsonneault, M. H.,
                     and Sarajedini, A. 1992}{388}{372}
\preprint{Christensen-Dalsgaard, J., Proffitt, C. R., and Thompson, M. J. 1993}
\ref\apj{Cox, A. N., Guzik, J. A., and Kidman, R. B. 1989}{342}{1187}
\ref\apj{Fontaine, G., and Michaud, G. 1979}{231}{826}
\refbook{\bysame 1979b, IAU Colloquium {\bf 59}, 192.}
\preprint{Guenther, D. B., Pinsonneault, M. H., and Bahcall, J. N. 1993}
\ref\apj{Iben, I., and MacDonald, J. 1985}{296}{540}
\ref\rmp{Ichimaru, S. 1982}{54}{1017}
\refbook Kippenhahn, R., and Weigert, A. 1991, {\it Stellar Structure
               and Evolution}
             (Springer-Verlag, Berlin).
\ref\apj{Michaud, G., Charland, Y., Vauclair, S., and Vauclair, G. 1976}
              {210}{447}
\preprint{Michaud, G. and Proffitt, C. R. 1992}
\refbook Michaud, G. and Vauclair, S. 1991,{\it Element Separation by
                 Atomic Diffusion}, in
                 Solar Interior and Atmosphere, ed. Cox, A.N., Livingston,
                     W.C. and Matthews, M.S. (University
                   of Arizona press, Tucson), p.312.
\ref\apjs{Montmerle, T., and Michaud, G. 1976}{31}{489}
\ref\aa{Noerdlinger, P. D. 1977}{57}{407}
\ref\apjs{\bysame 1978}{36}{259}
\ref\apjs{Paquette, C., Pelletier, C., Fontaine, G., and Michaud, G. 1986}
                   {61}{177}
\ref\apj{Pelletier, C., Fontaine, G., Wesemael, F., Michaud, G., and
                Wegner, G. 1986}{307}{242}
\refbook{Proffitt, C. R. 1993, {\it personal communication}; see also Paquette
e
 t al. 1986.}
\ref\apj{Proffitt, C. R. and Michaud, G. 1991}{380}{238}
\ref\annrev{Vauclair, S., and Vauclair, G. 1982}{20}{37}
\ref\aa{Vauclair, G., Vauclair, S., and Pamjatnikh, A. 1974}{31}{63}

\vfill\eject

%%%%%%%%%%%%%%%%%%%% Figures Captions %%%%%%%%%%%%%%

\centerline{\bf FIGURE CAPTIONS}
\bigskip

\noindent
FIG.~1.---
Variation of the hydrogen diffusion coefficients with the
hydrogen mass fraction in a pure hydrogen-helium plasma,
with $T=10^7\,{\rm K}$ and $\rho=100\,{\rm gcm}^{-3}$.
The solid lines represent the results obtained using Burgers
equations, with no approximation. The short-dashed lines represent
the results obtained when neglecting the heat fluxes. The
long-dashed lines are the results obtained by using a single
value for all the Coulomb logarithms, equal to 2.2. The dash-dot lines
are the results obtained when neglecting the heat fluxes
{\bf and} using a single
value for all the Coulomb logarithms, equal to 2.2.
(a) Pressure gradient coefficient $A_p$.
(b) Temperature gradient coefficient $A_T$.
(c) Hydrogen concentration gradient coefficient $A_H$.
(d) Relative errors due to the approximations. The short-dashed
line is the error on $A_p$ and $A_H$ when the heat fluxes are
neglected. The long-dashed lines are the errors
made by using $\ln\Lambda=2.2$, and the dash-dot line is the
error on $A_p$ and $A_H$ when both approximations are made.

\line{}
\noindent
FIG.~2.---
(a) Ratio of the exact hydrogen diffusion coefficients and those obtained
by various approximations, in terms of the hydrogen mass fraction X,
with $T=10^7\,{\rm K}$ and $\rho=100\,{\rm g cm}^{-3}$.
(a) Comparison with Bahcall and Loeb (1990).
(b) Comparison with Michaud and Proffitt (1992).
The solid lines and the dashed lines are the results obtained using
equations~(\lnmp) and equation~(\coullogii) respectively for the
Coulomb logarithms in the
Michaud and Proffitt formulae~(\apmp)-(\ahmp).

\line{}
\noindent
FIG.~3.---
Variation of the hydrogen diffusion coefficients with the hydrogen
mass fraction in a hydrogen-helium-oxygen plasma,
with $T=10^7\,{\rm K}$, $\rho=100\,{\rm g cm}^{-3}$, and
$Z=0.01$.

\line{}
\noindent
FIG.~4.---
Variation of the oxygen diffusion coefficients with the hydrogen
mass fraction in a hydrogen-helium-oxygen plasma,
with $T=10^7\,{\rm K}$, $\rho=100\,{\rm gcm}^{-3}$, and
$Z=0.01$.

\line{}
\noindent
FIG.~5---
Hydrogen diffusion coefficients in the present sun, as a function
of radius.

\line{}
\noindent
FIG.~6---
Contributions to the hydrogen diffusion velocity
in the sun due to each gradient.

\line{}
\noindent
FIG.~7---
Local diffusion time of hydrogen, oxygen, and iron
as a function of radius, in units of the age of the sun.

\line{}
\noindent
FIG.~8---
(a) Ratio of the exact hydrogen diffusion coefficients and those obtained
by Bahcall and Loeb (1990) as a function of the radius.
(b) Ratio of the exact hydrogen diffusion coefficients and those obtained
by Michaud and Proffitt (1992) as a function of the radius.

\line{}
\noindent
FIG~9---
Diffusion velocities in the contemporary sun. The solid lines are the exact
results; The short-dashed lines are the results of Bahcall and
Loeb (1990) (see eqs.~1-5 in BL); The long-dashed lines are the results
of Michaud and Proffitt (1992).
(a) Hydrogen diffusion velocity.
(b) Oxygen diffusion velocity.
(c) Iron diffusion velocity.

\line{}
\noindent
FIG.~10---
Coulomb logarithms in the sun.

\line{}
\noindent
FIG.~11---
Diffusion coefficients in the sun, as a function of radius.
The solid lines represent the results obtained using Burgers
equations, with no approximation. The dashed lines
represent
the results obtained using $\ln\Lambda=3.2$
for all the interactions.
(a) Pressure gradient coefficient $A_p$.
(b) Temperature gradient coefficient $A_T$.
(c) Hydrogen concentration gradient coefficient $A_H$.
(d) Relative error on the hydrogen and oxygen diffusion velocities
made by using $\ln\Lambda=3.2$.

\line{}
\noindent
FIG.~12---
Thermal coefficient for the electric field (see eq.~\Efield)
\end